\documentclass{sig-alternate}

\usepackage{epsfig}

\usepackage{balance}
\usepackage{caption}
\usepackage{color}
\usepackage{comment}
\DeclareCaptionType{copyrightbox}
\usepackage{epic}
\usepackage{epsf}
\usepackage{epsfig}
\usepackage{listings}
\usepackage{multirow}
\usepackage{textcomp}
\usepackage{times}
\usepackage{url}
\usepackage{xspace}    
\usepackage{latexsym}
\usepackage[subrefformat=parens,labelformat=parens]{subfig}
\usepackage{caption}
\usepackage{afterpage}
\usepackage{amstext}   
\usepackage{amsmath}
\usepackage{url}                                                           
\usepackage[pdftex,colorlinks=true,citecolor=black,filecolor=black,%
            linkcolor=black,urlcolor=black]{hyperref}
\usepackage{graphicx}
\usepackage{tabularx}
\usepackage[table]{xcolor}
\usepackage{soul}

\makeatletter
\def\@copyrightspace{\relax}
\makeatother

\makeatletter
\newcommand{\labitem}[2]{%
\def\@itemlabel{\textbf{#1}}
\item
\def\@currentlabel{#1}\label{#2}}
\makeatother

\newcommand{\Sys}{{\tt DAP}\xspace}
\newcommand{\Cost}{5.1 ms\xspace}
\newcommand{\Overhead}{17\%\xspace}

\usepackage{todonotes}
\newcommand{\nabilsays}[1]{\todo[inline, color=red]{Nabil: #1}}

\makeatletter
\def\blfootnote{\xdef\@thefnmark{}\@footnotetext}
\makeatother

\makeatletter
\newcommand\footnoteref[1]{\protected@xdef\@thefnmark{\ref{#1}}\@footnotemark}
\makeatother

\newif\ifanon
\anontrue
\anonfalse

\ifanon
\makeatletter
\def\@maketitle{\newpage
 \null
 \setbox\@acmtitlebox\vbox{%
\baselineskip 20pt
\vskip 1em                   
   \begin{center}
    {\ttlfnt \@title\par}       
    \vskip -1em                
{\subttlfnt \the\subtitletext\par}\vskip .25em
    {\baselineskip 26pt\aufnt   
     \lineskip 0em             
     \begin{tabular}[t]{c}\@author
     \end{tabular}\par}
    \vskip -2em               
   \end{center}}
 \dimen0=\ht\@acmtitlebox
 \unvbox\@acmtitlebox
 \ifdim\dimen0<0.0pt\relax\vskip-\dimen0\fi}
\fi

\title{Retrofitting Applications with Provenance-Based Security Monitoring
\ifanon
\else
\titlenote{This work is sponsored by 
the Assistant Secretary of Defense for Research \& Engineering under Air Force Contract 
\#FA8721-05-C-0002. Opinions, interpretations, conclusions and recommendations are those of the 
author and are not necessarily endorsed by the United States Government.}
\fi}

\ifanon
\numberofauthors{1} 
\author{
\alignauthor
}
\else
\numberofauthors{3}
\author{
\alignauthor
Adam Bates\\
   \affaddr{University of Illinois at Urbana-Champaign} \\
   \affaddr{batesa@illinois.edu} \\
\alignauthor
Kevin Butler, Alin Dobra, Brad Reaves\\
   \affaddr{University of Florida} \\
   \affaddr{\{butler,adobra,reaves\}@ufl.edu} \\
\alignauthor
Patrick Cable, Thomas Moyer, Nabil Schear\\
   \affaddr{MIT Lincoln Laboratory} \\
   \affaddr{\{cable,nabil,tmoyer\}@ll.mit.edu} \\
}
\fi

\begin{document}


\date{}

\maketitle

\begin{abstract}
Data provenance is a valuable tool for detecting and preventing cyber attack, providing insight into the nature of suspicious events.
For example, an administrator can use provenance to identify the perpetrator of a data leak, track an attacker's actions following an intrusion, or even control the flow of outbound data within an organization.
Unfortunately, providing relevant data provenance for complex, heterogenous software deployments is challenging, requiring both the tedious instrumentation of many application components as well as a unified architecture for aggregating information between components.

In this work, we present a composition of techniques for bringing affordable and holistic provenance capabilities to complex application workflows, with particular consideration for the exemplar domain of web services. We present \Sys, a transparent architecture for capturing detailed data provenance for web service components.
Our approach leverages a key insight that minimal knowledge of open protocols can be leveraged to extract precise and efficient provenance information by interposing on application components' communications, 
granting \Sys compatibility with existing web services without requiring instrumentation or developer cooperation.
We show how our system can be used in real time to monitor system intrusions or detect data exfiltration attacks while imposing less than \Cost end-to-end overhead on web requests. Through the introduction of a garbage collection optimization, \Sys is able to monitor system activity without suffering from excessive storage overhead. 
\Sys thus serves not only as a provenance-aware web framework, but as a case study in the non-invasive deployment of provenance capabilities for complex applications workflows.

\end{abstract}

\section{Introduction}
\label{sec:intro}
Data provenance describes the history of the execution of computing systems, providing detailed explanations as to how data objects were created and came to arrive at their present state. Traditionally, data provenance has
been extremely valuable to performing forensics following an attack \cite{bbh+2014,lzx2013,zfn+2011}.  For example,
provenance can indicate which hosts, processes, files, and data have been affected during the attack and cue cleanup and recovery \cite{tbg+2011}.
Additionally, provenance is of value in virtually any circumstance where a context-sensitive decision must be made about a piece of data. 
Provenance-aware solutions have been proposed for access controls \cite{nps12,pds12}, data leakage \cite{jsd+2011}, malware detection \cite{gbm+2010}, scientific processing \cite{abj2006}, and distributed computing \cite{gm2011}.

Unfortunately, in many complex applications where data provenance would be of greatest value, there is not a general solution for provenance deployment.
Although a variety of tools to aid in the design of provenance-aware applications have been proposed \cite{hsw2009,lzx2013,ms2012,zfn+2011}, modern software is created through the composition of many software artifacts that were written by different developers.
Provenance-aware operating systems \cite{btb+2015,gl2007,gt2012,mkb+2009,pmm+2012,sc2005} provide an alternative to ad hoc instrumentation efforts, but are not a complete solution in practice due to semantic gap problems.
For example, an operator may want to use provenance to ensure PCI compliance on a credit card database; however, the abstraction level with which the operator wants to work (i.e., credit card records in a database) does not match the system level objects over which provenance is captured (i.e., processes, files, pipes, etc.).
This semantic gap also gives rise to the problem  of {\it dependency explosion} -- in a long-running program, each output must conservatively be assumed to have derived from all prior inputs \cite{lzx2013}. 
While existing approaches to application layer provenance may overcome one instance of the dependency explosion problem \cite{lzx2013,Lee-ccs13}, they are not a panacea to this semantic gap problem, as they cannot observe the semantics all components in a complex workflow.

In this work, we introduce a low-cost methodology for retrofitting application workflows with provenance capabilities through a composition of different introspection methods.
We present the design and implementation of a unified provenance-aware architecture that includes both novel workflow reconstruction techniques as well as other known approaches to provenance collection.
Our exemplar provenance-aware workflow mechanism, \Sys,\footnote{Dapping is a form of fly fishing that causes  minimal disturbance to the water. Likewise, our \Sys system is minimally invasive to application workflows while extracting precise contextual metadata.}
is designed with consideration for the unique challenges and opportunities presented by web service environments. 
\Sys is a transparent collection agent that captures detailed provenance of application workflows without suffering from the semantic gap problems of system-level collection.
Our approach leverages minimal knowledge about common workflow structures in order to extract precise and efficient provenance.
In particular, \Sys leverages the widespread adoption of the SQL syntax, transparently interposing on database connections to interpret and extract the provenance of database transactions.
\Sys is compatible with a large percentage of existing web services, generating concise and understandable provenance with little or no system modification.

Our contributions can be summarized as follows:

\begin{itemize}

\item We present the design and implementation of \Sys, a minimally invasive, low overhead framework for capturing workflow provenance. \Sys combines state-of-the-art provenance techniques with novel components that aggregate the provenance of application objects. We also address the challenging problem of integrating provenance from different capture points under a common namespace.

\item We present an extended case study through which we demonstrate \Sys to be an effective means of reasoning about, detecting, and actively preventing Internet-based attacks. In particular, we demonstrate that our system can be used as a means of preventing SQL Injection (SQLi)-based data exfiltration, one of the most widespread and insidious threats to the Internet today. We also show how \Sys can be used in concert with other technologies to track system layer attacks against the web server.

\item In evaluation, we show that our implementation imposes just \Cost of overhead on web application requests, and microbenchmark individual steps in our system to arrive at a better understanding of this cost. Automatic provenance collection is known to impose excessive storage overheads on long running systems;  however, through applying a garbage collection optimization, we show that our mechanism can monitor for active SQLi attempts while maintaining sublinear growth in storage burden.

\end{itemize}

\section{Web Application Provenance}
\label{sec:background}
\begin{figure}[t]
\centering
\begin{minipage}[b]{1.0\linewidth}
\includegraphics[width=1\linewidth]{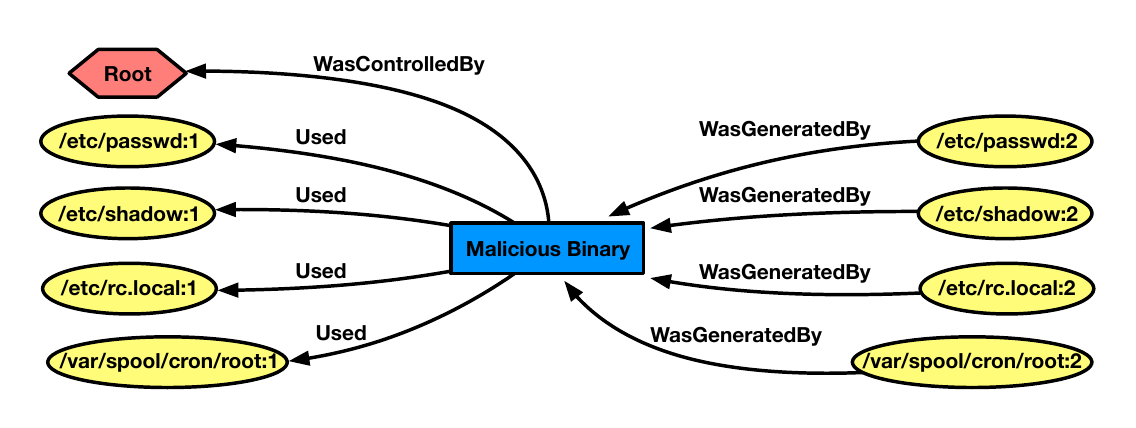}
\end{minipage} 
\caption{A provenance graph showing the actions of a potentially malicious binary that is running with root privileges. Edges encode relationships that flow backwards into the history of system execution, and writing to an object creates a second node with an incremented version number.}
\label{fig:attack_example}
\end{figure}
 
Data provenance describes the actions taken on a data object from genesis onwards, including how it came to exist in its present state. Provenance can be queried to answer questions such as ``{\it What datasets were used in the creation of this data object?}" and ``{\it In what environment was this data object produced?}"   A standard representation for data provenance is a directed acyclic graph which is specified in the W3C PROV data model \cite{w3c}, which we will use throughout this work. 
An example provenance graph plotting the execution of a  potentially malicious binary is shown in Figure \ref{fig:attack_example}.
This binary, while running with root privileges, first read several system files, including {\tt /etc/shadow} and {\tt /etc/rc.local}. It then wrote to those files in an attempt to gain persistent access to the system.
In the graph, edges represent relationships between different system objects.
To prevent cycles from forming in the graph, writing to an object triggers the creation of a new node that represents a new version of the object. 

\begin{figure}[t]
\centering
\includegraphics[width=.75\linewidth]{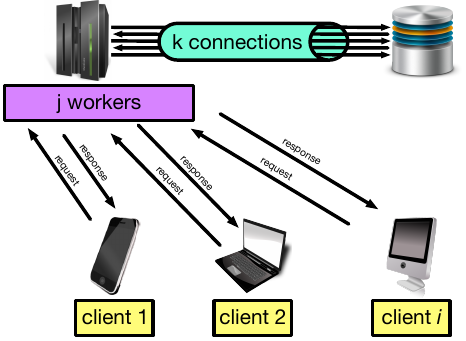}
\caption{Diagram of a web service architecture. To accurately track provenance, it is necessary to track individual client requests from the network, through the server, to the database, and back.}
\label{fig:webservice}
\end{figure}

Web applications present a challenging scenario for provenance because of their heterogenous nature; web requests traverse the operating system, web server, web application, and database management system, each of which maintain their own internal semantics. Consider the representative scenario in Figure \ref{fig:webservice}. There are $i$ different Internet clients sending requests.   Network requests pass from the the operating system to the web server software, which the server handles concurrently with $j$ different workers. The workers' database transactions are then multiplexed between $k$ different connections. Unfortunately, there is not necessarily any equivalence between the numbers $i$, $j$, and $k$. 

To reason about the attack service of a web application it is necessary to understand the actions of each of these components, yet deploying provenance capabilities in this domain is particularly challenging due to the complexity of this workflow. 
Provenance-aware operating systems such as PASS \cite{mhb+2006} , LPM \cite{btb+2015}, and ProTracer \cite{mzx2016} provide a single point of observation for all system activity, but are not a complete solution due to the semantic gap that divides the system and application layers.
For instance, these systems would struggle to disambiguate web server requests as autonomous units of work (i.e., {\it dependency explosion} \cite{lzx2013}), leading to the false conclusion that each server response was dependent on all previous client requests. 
Lee et al.'s LogGC \cite{Lee-ccs13} and BEEP \cite{lzx2013} system provide space-efficient forensics for application monitoring, but suffer from the same semantic gap problem as provenance-aware operating systems due to their reliance on system audit logs.

The remaining alternative to the above approaches is to undertake a tedious instrumentation effort of the web application.
Although provenance libraries exist that simplify the manual instrumentation of source code \cite{ms2012}, this approach requires additional resources and domain-specific knowledge that is unlikely to be available to most web developers.
Furthermore, instrumentation could extend past the primary software artifact to its dependencies, including the web server, runtime framework, and other third party libraries.
This solution may even require re-architecting the web service use a provenance-aware database management services such as Trio \cite{w2004}, DBNotes \cite{ctv2005}, and ORCHESTRA \cite{kgi+2013}.
Due to the extraordinary capital required by this approach, we conclude that is not a viable solution to creating provenance-aware web services.
What is needed instead is a means of retrofitting provenance into existing services with minimal cost to web developers.

\section{Design}
\label{sec:design}
\begin{figure*}[t]
\centering
\includegraphics[width=1\linewidth]{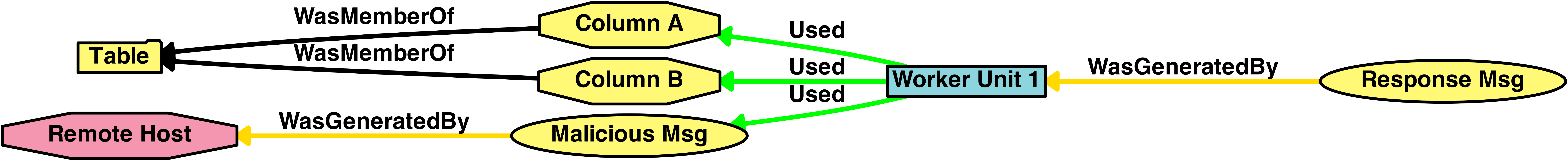}
\caption{The provenance graph of a typical web session that we wish to observe with our architecture.}
\label{fig:dbprov_mockup}
\end{figure*}

\subsection{Threat Model \& Assumptions}
\label{sub:threat}

The attack surface we consider in this work is that of a typical web application. 
By connecting to the application's external listening ports the attacker may attempt a variety of misdeeds on the system. 
The attacker may attempt to exfiltrate data from the web application through iterative command injection (e.g., SQLi) attacks.
A successful exfiltration attack will involve repeated command injections as the attacker attempts to discover the location of valuable data.
The attacker may also attempt to use command injection to inject spurious data into the database for the purposes of privilege escalation or cross-site scripting.
Alternately, the adversary's target may not be the web application but the system itself.
The attacker may be attacking the web server in order to compromise other services on the host or to move laterally to other hosts on the network \cite{zpj+2006}.

We make the following assumptions about the security of each web service component. We conservatively assume that the web server, web application, and database engine are all subject to compromise. 
These components may begin to lie about their actions on the system at any time, but we assume that at least one provenance record of the attacker's access attempt is recorded prior to compromise. 
We also assume that the integrity of the host kernel is assured. 
This condition is made more reasonable through the deployment of kernel hardening techniques, integrity measurement, and mandatory access control (e.g., SELinux) that protects the operating system's trusted computing base. 
Finally, we assume that the novel components our system introduces are {\it not} subject to compromise. As we will later show, these mechanisms are small and simple enough to be subjected to rigorous audit, and can also be protected through system hardening techniques.

\subsection{System Goals}
\label{sub:goals}

\begin{itemize}

\labitem{G1}{itm:g1} {\bf Complete.} Our system must offer a complete description of individual requests as they pass through an application workflow. The record must remain complete in the presence of unexpected events triggered by attacker behavior, such as command injection attacks or binary exploitation. If we elect to forego provenance capture at a given system component, the recorded provenance must provide sufficient context to reconstitute the entire workflow.

\labitem{G2}{itm:g2} {\bf Integrated.} Our system must combine provenance from different operational layers  in a salient manner that provides a coherent explanation of application activity to the administrator. Provenance generated by different capture agents must share a common namespace, and each capture agent must be able to accurately reference the activities of other agents. 

\labitem{G3}{itm:g3} {\bf Minimally Invasive.} Provenance, like security, is often perceived as a cost burden. Our system must therefore impose a bare minimum number of modifications to existing system components, including the application and backend infrastructure (e.g. database engine, web server). Optimally, our solution would not make any changes to  existing software, instead introducing independent mechanisms  so that the system would continue to function correctly as software in the application is periodically upgraded.

\labitem{G4}{itm:g4} {\bf Widely Applicable.} To further advocate for the deployability of provenance-aware applications, our efforts in the development of the system should not be limited to the benefit of a particular application, backend component, or architecture. Instead, our system should be immediately compatible with a broad number of existing applications.

\labitem{G5}{itm:g5} {\bf Defensive Capabilities.} While provenance is invaluable to forensic investigation after an attack has occurred, attacks on Internet domains are frequent and relentless. Therefore, our system must be fast enough to provide real-time assistance to the defense of the host. This includes the ability to detect and explain attacks as they occur and aid in system recovery in the event of a successful attack.

\end{itemize}

\subsubsection{Provenance Definition}
\label{sub:define}

As provenance is codified in dramatically different ways throughout the literature, from exhaustive descriptions of system activity \cite{pmm+2012} to lightweight proofs of program execution \cite{lm2010}, an important step in the design of our system is to identify the scope and granularity of the events we wish to observe. As the ultimate goal of our system is to observe the attacker described in Section \ref{sub:threat}, the provenance we collect must exhaustively describe the manner in which attacker inputs interact with the application workflow. We illustrate this with a typical web application, where we must be able to track a client request from receipt on the host, through a specific worker in the web server, through a database request and response, until a response is crafted by the web application and returned to the remote client. We must also be able to differentiate between different client requests, even if they are performed by the same worker or re-use the same database connection.

In order to satisfy Goals \ref{itm:g3} and \ref{itm:g4}, our system must avoid modifying the database management system. As a consequence of this, we will be unable to know the precise records that are impacted by a query. Therefore, we must describe SQL objects not at the granularity of database records, but instead as database columns and tables, which can be inferred from the query itself. As we will show in Section \ref{sec:analysis}, this coarser granularity is well suited to explaining command injection techniques, which often involve access to columns that should never be returned to the user. 

With this in mind, the goal of \Sys is to produce provenance explanations for individual web requests like the one shown in Figure \ref{fig:dbprov_mockup}. Network activity is tracked at the system granularity, the web application is tracked at the granularity of individual units of work, and database objects are tracked as columns and tables. Provenance captured at different sources will be integrated through their shared relation to the web application worker during a given unit of work.

\begin{figure}[t]
\centering
\includegraphics[width=1\linewidth]{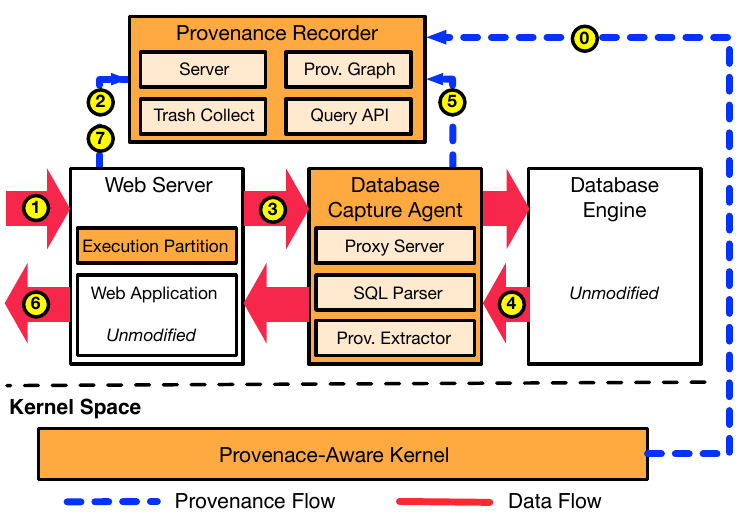}
\caption{Overview of the \Sys architecture. Provenance-Aware components are shaded in orange. No changes are required to the Database Engine or Web Application; instead,  provenance is generated by interposing on the connection between the Web Application and Database Engine. A small change to the Web Server is required to facilitate execution partitioning.}
\label{fig:dbprov_architecture}
\end{figure}

\subsection{Provenance Capture: Overview}
\label{sub:overview}

In order to achieve the above goals, what is needed is a comprehensive provenance architecture that is able to understand the semantics of the web server, database, and operating system in unison. 
An overview of our solution to this challenge, \Sys, is shown in Figure \ref{fig:dbprov_architecture}. 
Provenance-aware components are shaded in orange. 
\Sys introduces several provenance-aware components, but requires no modification to the Web Application or Database Engine. 
The provenance-aware components are a small and re-usable modification to the Web Server to facilitate execution partitioning, a Database Capture Agent that transparently proxies all traffic between the Web Application and the Database Engine, and a Provenance Recorder that aggregates provenance information between the different parties.
Our system also assumes that system layer provenance is being collected, which can be obtained through use of a custom provenance-aware kernel \cite{btb+2015,mkb+2009,pmm+2012,sc2005} or user-space system monitor \cite{gl2007,gt2012}.

The workflow for provenance collection is as follows: (1) a remote host makes a request to the Web Application; (2) a small modification to the Web Server performs {\it execution partitioning} \cite{lzx2013}, notifying the Provenance Recorder whenever the Web Application has started a new autonomous unit of work; (3) the Database Capture Agent proxies and subsequently parses a query issued to the Database Engine; (4) after measuring the impact of the query by parsing the Database Engine response, the Database Capture Agent (5) transmits provenance information to the Provenance Recorder; (6) as the Web Applications transmits a response to the remote host, (7) the Web Server notifies the Provenance Recorder that the unit of work has ended; throughout execution, (0) the provenance-aware kernel generates provenance for all activities that are not being explicitly disclosed by the Web Server or Database Capture Agent.

In the remainder of this section, we will describe in greater depth the operation of the Database Capture Agent as well as introduce a provenance-based defensive mechanism. The Execution Partition and Provenance-Aware Kernel components rely on known techniques and are discussed at greater length in Section \ref{sec:implementation}. 

\subsection{Provenance Capture: Database}
\label{sub:proxy}

A fundamental design consideration in our system was the manner in which \Sys would observe communication between the Web Application and Database Engine. One possibility would be to instrument the web service to extract database queries. This would have made our solution application-specific, violating Goal \ref{itm:g4}. Another possibility would be to instrument the database, or to use an existing provenance-aware database. However, instrumenting a database engine would also limit our solution to the benefit of a particular database service, violating Goal \ref{itm:g4}. In turn, using a provenance-aware database would require re-architecting the web service, violating Goal \ref{itm:g3}.

Instead, we chose to implement an explicit TCP proxy that interposes on communications between the application and database.
The only change required by this approach is that either the application or the database change the port over which they communicate with one another, allowing the proxy to interpose.  
These types of configuration options are nearly always exposed and easily modifiable in both database and web application software.
We rejected a fully transparent solution that used \texttt{iptables} to capture packets between the components, as it would substantially increase the complexity of the capture agent.
We also rejected achieving interposition by modifying an existing database connection library as this would limit the applicability of our agent to applications with that particular dependency, violating Goal \ref{itm:g4}.

\subsubsection{Query Parsing}
\label{sub:parse}
\begin{figure}[t]
\centering
\includegraphics[width=1\linewidth]{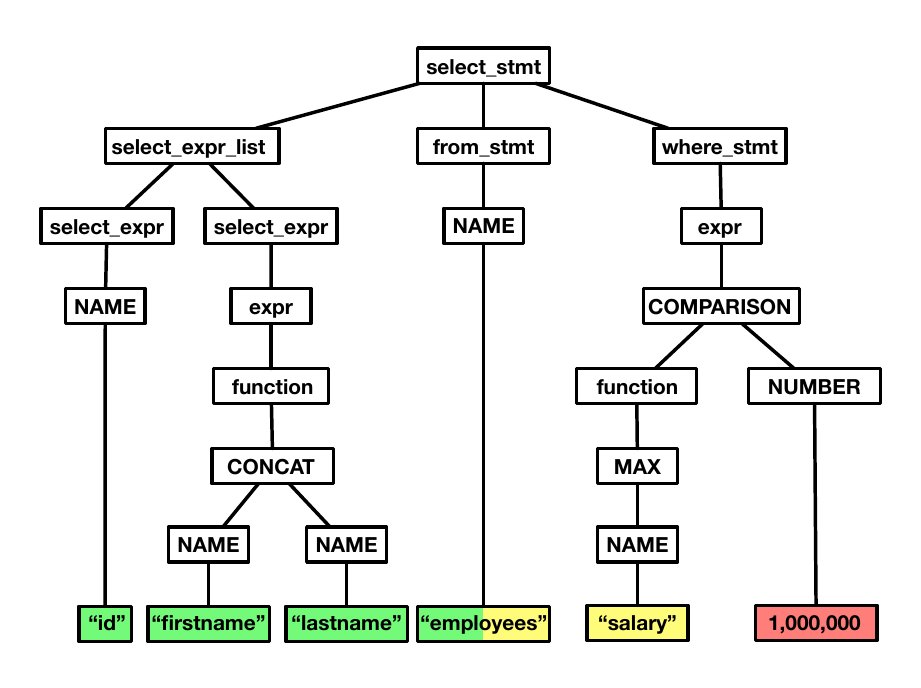}
\caption{A simplified example of a SQL parse tree for the statement ``{\it SELECT employee\_id, CONCAT(firstname, lastname) FROM employees WHERE MAX(salary) $>$ 1,000,000}''. The leaves of the tree are color-coded by their provenance extraction decision condition. Columns in the select expression are noted by a {\tt SQL\_READ} provenance event. Columns present in other subexpressions are noted by a {\tt SQL\_USED} event. Non-persistent entities such as numbers, and functions are not considered for provenance extraction.}
\label{fig:parsetree}
\end{figure}

After proxying the web application's traffic, we make use of a query grammar and parser to extract fine-grained provenance information from database queries. We chose to focus on the SQL language in this work due to its widespread use. While in reality SQL comes in many different flavors and varies by database management system, our needs are foundational enough that the syntactical differences between SQL variants can be largely ignored.

The output of a SQL parser can be visualized as a parse tree, a simplified example of which is shown in Figure \ref{fig:parsetree}. This tree is is a SELECT statement that contains a FROM clause (required) and a WHERE clause, which is  one of several optional clauses. We use this example to demonstrate how \Sys handles the various data objects contained in a SQL query:

\noindent
{\bf Data Accessed:} We refer to the named objects referenced in the primary clause of the query as accessed data. These are the objects that will be returned by the database in its response to the query. When a query is parsed, \Sys generates a {\tt SQL\_READ} provenance event for each piece of accessed data. The accessed data in Figure \ref{fig:parsetree} are the {\it employee\_id}, {\it firstname}, and {\it lastname} columns, and the {\it employees} table.

\noindent
{\bf Data Referenced:} Named objects that appear in subsequent clauses of the query are not explicitly returned by the database, but nonetheless inform the response message.  Consider again the example query in Figure \ref{fig:parsetree} -- while employee salaries are not returned in the query, the response implicitly informs the querier of which employees salaries are greater than \$1,000,000. Referenced data therefore represents a dangerous side channel for information leakage. However, in some environments it may be unnecessarily conservative to treat all referenced data as accessed data. To account for this, we introduce a {\tt SQL\_USED} event to describe referenced data. 

\noindent
{\bf Ephemeral Data:} Query expressions also include non-persistent data objects, such as numbers and string literals. In the case of SELECT statements, ephemeral data can manipulate the records and values returned by a query, but not the columns accessed and referenced. We choose to ignore ephemeral data for the case of provenance extraction. 

During parsing, the SQL grammar tracks named and referenced data via synthesized attributes. The name, and also the prefix if present, is added to a linked list as the statement is parsed. At the root of the statement, a function determines the appropriate prefix for each column given the tables used in the FROM clause.

While we use the SELECT statement in the scenario above, the same rules can be applied to other statements.
SHOW and DESCRIBE statements can be treated the same way as SELECT statements. 
For expressions that write to the database, we introduce the provenance event {\tt SQL\_WASGENERATEBY}.
This event is used to describe the column and table references that appear in the primary clauses of INSERT and UPDATE expressions.
This event contains the same fields as the {\tt SQL\_READ} event, but is handled differently by the Provenance Recorder.
When a {\tt SQL\_WASGENERATEBY} is received, the Recorder will create a new node for the accessed object with an incremented version number.
Subsequent {\tt SQL\_READ} and {\tt SQL\_USED} events will be linked to the newer node in order to prevent cycles from forming in the graph.
While we do not explicitly address any other statement types in this work, our rules generalize to any expression that reads from or writes to the database.

\subsubsection{Parsing Challenges}
\label{subsub:challenges}

When extracting provenance from non-trivial SQL statements, a variety of challenges arise. We came across a number of such challenges while designing and implementing \Sys. We describe our solutions to each problem below:

\noindent
{\bf Parsing Challenge \#1: Wildcards}. Through use of the wildcard character, SQL statements are able to reference all columns in a table without explicitly naming them. To address this, we provide the Provenance Recorder a schema description, which allows it to translate the wildcard character into the associated columns for the given table. In our implementation, we obtain the schema through use of the {\tt mysqldump} command.

\noindent
{\bf Parsing Challenge \#2: Aliases}. Any value in a SQL statement can be aliased to another name. The challenge in resolving aliases is that an alias may be referenced in one clause of the query, but defined in another. 
To address this problem, our SQL grammar makes use of synthesized attributes to track references and definitions of aliases. At the top level of the parse tree, the list of referenced aliases are then resolved to their true table and column names. In effect, this means that \Sys unaliases named objects during parsing, ensuring that the extracted provenance is unobfuscated.

\noindent
{\bf Parsing Challenge \#3: Nested Queries}. An additional obstacle we faced in the design of our grammar was that of nested queries. In SQL, full statements can be indefinitely nested within one another. For example, ``SELECT A FROM (SELECT id AS A FROM employees)'' is a valid statement. Nested queries can be used to further obfuscate the true origin of a data object. Our solution to this is to modify the synthesized attribute routines described above. At the root of each subquery in the parse tree, objects in the subquery are unaliased, and the named and referenced objects used by the subquery are transferred to the parent query. Additionally, the alias mapping is passed to the parent query. This allows \Sys to unnest queries as the statement is parsed.

\begin{figure}[t]
\centering
\includegraphics[width=1\linewidth]{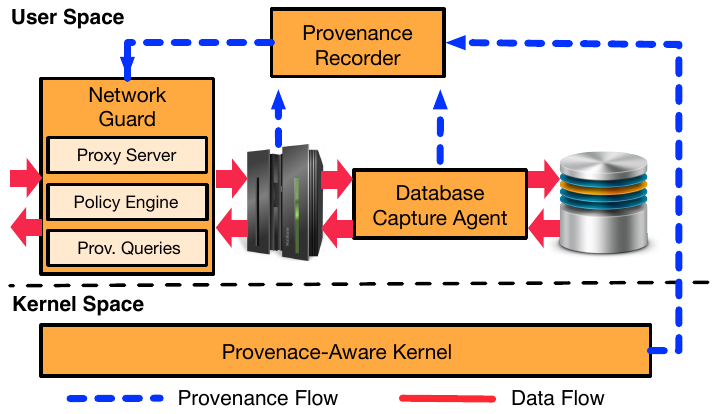}
\caption{We extend \Sys with analysis components that provide real-time detection of data exfiltration attacks (Network Guard).}
\label{fig:network_guard}
\end{figure}

\subsection{Provenance-Based Security Mechanism}
\label{sub:networkguard}
Given the above provenance capture agents, \Sys can be used to thwart SQLi-based data exfiltration attacks in real-time by performing provenance-based whitelisting of outbound server responses. We accomplish this through introducing a Network Guard component to the  \Sys architecture as shown in Figure \ref{fig:network_guard}. The Network Guard is another TCP proxy server placed between web server and the network.  It takes no action on incoming traffic, but inspects outbound network traffic from the server. When the guard intercepts a transmission to the remote host, it examines the network context to identify the process id of the worker thread. It then issues a query to the Provenance Recorder to obtain the list of data ancestors used by the worker during the current unit of work. 

In order for the Network Guard to prevent exfiltration, we require a means of encoding developer or administrator intent about web service workflows. To do so, we introduce a simple policy language. The Network Guard accepts a list of policy rules and checks the list of ancestors against each rule before permitting the message to be transmitted. Rules take the form: 

\begin{center}
$< RULE, [ Tab_1.Col_1, \dots, Tab_n.Col_n], SIZE>$
\end{center}

$RULE$ can be set to $ALLOW$ to whitelist ancestors, or $DENY$ to blacklist ancestors.
A list of SQL objects can then be specified to match against in the response messages ancestry. We support lists of objects, rather requiring multiple policy entries for each object in the list, because at times the fusion of data objects reaches a different sensitivity level than the level of each individual object (e.g., Personally Identifiable Information\footnote{\label{pii} See NIST SP 800-122}). $SIZE$, expressed in bytes, approximates the impact of a query.  It can be used to limit the amount of SQL data that should be returned to a client in a response message, or can be disabled by setting to 0.

\subsection{Security Analysis}
\label{sec:analysis}

\noindent
Our arguments for the security of \Sys are as follows:

\noindent
{\bf Complete (\ref{itm:g1}).} \Sys is able to differentiate between individual web requests through a minimal modification to the Apache 2 server that performs execution partitioning. Transactions with the database are observed by \Sys's Database Capture Agent. Because the port for the database engine is hard-coded into the web application, there is no other means to reach the database then to go through the Capture Agent. In the event that the server is compromised, \Sys can continue to track the actions of the attacker on the system through use of the provenance-aware kernel, whose trusted computing base can be isolated from the rest of user space using SELinux enforcement \cite{btb+2015}.


\noindent
{\bf Integrated (\ref{itm:g2}).} A major challenge in layered provenance systems is establishing a uniform namespace for objects \cite{secondprovenancechallenge}. In \Sys, we integrate provenance from different capture agents through tracking process ids (pid). In our implementation, the web server is configured to run in pre-fork mode in which all workers receive a unique pid. When a request is proxied by the Database Capture Agent, the worker's pid is recovered through matching the network context in the {\tt sockaddr\_un} struct against the list of active tcp sockets returned by {\tt netstat}. The pid is then embedded in all subsequent provenance events generated by the Database Capture Agent. To integrate \Sys provenance with system-layer provenance, we introduce a {\tt pid\_to\_provenance} syscall that accepts a pid and returns the universally unique identifier associated with the provenance of the process' fork in the kernel. Thus, all layers share a common language to describe an activity.

\noindent
{\bf Minimally Invasive (\ref{itm:g3}).} { \Sys works without requiring any changes to existing web applications.} In our implementation, the web server is minimally modified through the introduction of a new header file as well as the insertion of 3 lines of code in the existing source. The database requires no change except to be reconfigured to listen on another port, which can be done without recompilation. If the ability to track attacker actions after a server compromise is desired, a provenance-aware kernel must be installed on the machine. However, \Sys is able to track service layer attacks, such as command injection, without this capability.

\noindent
{\bf Widely Applicable (\ref{itm:g4}).} We confirmed that our \Sys implementation is compatible with the Apache 2 and Tomcat (via mod\_jk) web frameworks, as well as with the MySQL and PostGreSQL database engines. \Sys's~use of the SQL grammar is simple enough that it should work with any SQL variant with only minimal modification. In Section \ref{sec:disc}, we discuss whether our approach generalizes to NoSQL databases.

\noindent
{\bf Defensive Capabilities (\ref{itm:g5}).}
 The Network Guard component proxies outbound network traffic and searches its provenance for evidence of data exfiltration.  
 An administrator can use this tool to prevent certain data objects from ever being returned to the client, or even to terminate connections if a suspicious amount of data is being transmitted to the client.
Moreover, we reduce the attack surface of \Sys by isolating the complex parsing procedure from its parsing responsibilities, making it more difficult to disable.

\section{Implementation}
\label{sec:implementation}

We have implemented \Sys for the Linux operating system.
We have tested that our system works with both MySQL and PostGreSQL by using their command line clients.
In both cases, the only change required to the database engine was to modify the port on which they listen for connections.
We also confirmed that \Sys worked correctly with a variety of web applications and tools including MediaWiki\footnote{\label{mediawiki} Available at \url{https://www.mediawiki.org}.} (PHP-based), UnixODBC\footnote{\label{unixodbc} Available at \url{http://www.unixodbc.org}.}, and a suite of Tomcat-based applications released with the AMNESIA \cite{ho2005} evaluation testbed\footnote{\label{amnesia_testbed} Available at \url{http://www-bcf.usc.edu/~halfond/testbed.html}.}. 

\subsection{Provenance-Enhanced Web Server}

We have instrumented the Apache 2 web server (version 2.2.31) to perform execution partitioning on each TCP socket.
The server is configured to run in pre-forked mode\footnote{\label{prefork} See \url{http://httpd.apache.org/docs/2.2/mod/prefork.html}.}.
Because we did not have access to the BEEP tool, we manually instrumented the source code.
To minimize the impact on the rest of the Apache 2 code base, all of the logic required to report execution partitions to the Provenance Recorder was included in a single header file.
As a result, we inserted just 3 lines of code into the existing source files. The changes were made to {\tt server/config.c}. The first insertion included our header file. The second two lines are placed before and after the call to {\tt ap\_run\_handler} in the {\tt ap\_invoke\_handler} function:

{\footnotesize
\begin{verbatim}
/* DAP -- Transmit "Unit Start" Message here! */
char * uuid = dap_unit_start(
                  r->connection->remote_addr);

/* Handle the request */
result = ap_run_handler(r);

/* DAP -- Transmit "Unit End" Message here! */
dap_unit_end(uuid);
\end{verbatim}
}

The {\tt dap\_unit\_start} function generates a UUID to associate with the unit of work, then transmits the {\tt UNIT\_START} message to the Provenance Recorder that contains the UUID and the {\tt remote\_addr} struct. The {\tt dap\_unit\_end} function transmits a {\tt UNIT\_END} message to the Provenance Recorder that contains the UUID, then frees the UUID character array.

Instrumenting this layer of the Apache 2 stack offers several advantages. First, it resides beneath the various error and security filters performed by the server, ensuring that we do not generate provenance for requests that are later rejected. Second, it resides above the file-specific handler module, so we are able to instrument both static web pages and dynamic web applications. As a result, our provenance-enhanced Apache 2 works for various handler modules, not just HTTP. For instance, when we later recompiled Apache with the {\tt mod\_jk} module\footnote{\label{mod_jk} See \url{http://tomcat.apache.org/connectors-doc}.}, 
we found that our code also worked on Tomcat applications without any modification.

\subsection{Database Capture Agent}

Our capture agent is a multithreaded TCP proxy server that listens on the database engine's assigned port. Once connected, the server extracts database queries issued by the web application. It then passes them through a Bison parser that makes use of a publicly available SQL grammar\footnote{\label{sqlparser} Available at \url{https://github.com/hoterran/sqlparser}.}. We extended the grammar to aggregate the columns and tables accessed by the query as described in Section \ref{sub:parse}.

When a new connection is proxied, the Capture Agent first recovers the process ID (pid) of the sender by matching the network context in the {\tt sockaddr\_un} struct against the list of active tcp sockets returned by {\tt netstat}. 
After parsing the query, the Capture Agent inspects the list of database objects accessed.
It then creates a provenance event for each object, which is a tuple of the form $<pid,~relationship,~column,~table>$.
$Relationship$ is one of the several relationships specified in Section \ref{sub:proxy}, and are also consistent with the W3C PROV-DM model.
Once the new provenance event is created, it is sent to the recorder through a Unix socket.
Rather than design our own protocol, we extend the space-efficient Hi-Fi protocol \cite{pmm+2012} to support several new kinds of events.

In order to ensure application stability, the Database Capture Agent's parsing functionality is isolated from the TCP proxy server. The parser is implemented as a separate binary that is invoked by the proxy using the {\tt system} syscall. After running the parser, the proxy checks its exit code status. If the parser exited in a bad state, indicating a potential attack, the proxy drops the connection and transmits the input to the Provenance Recorder as an attribute to be added to the provenance graph. Because the proxy process does not directly inspect messages from the application, we are confident that it cannot be disabled by malcrafted inputs.

\subsection{Provenance-Aware Kernel}

In the event that the Web Server is compromised, the above components alone are insufficient to track attacker actions. This is because the attacker will no longer be limited to the Web Application workflow, but will instead be able to take any action on the system with the privileges of the Web Server. More dangerously, the attacker may be able to use this foothold to escalate to root privileges. Notably, all application-based provenance-tracking \cite{lzx2013} suffers from the same limitation of being unable to reliably record provenance after the point of compromise. To ensure the ability to track attacker actions after a server exploit, we ran our application on top of a provenance-aware operating system.  We made use of the Linux Provenance Modules project for our provenance-aware kernel. We configured LPM to make use of the Hi-Fi module \cite{pmm+2012}, and deployed the system as described in \cite{btb+2015}. 

\subsection{Provenance Recorder}

The Provenance Recorder is responsible for aggregating provenance from the different capture agents and representing it in an efficiently queried in-memory graph.
We implemented the Recorder in C++ using the SNAP graph library.
The Recorder listened for new provenance events over Unix sockets.
Different provenance events are handled as described in Section \ref{sub:proxy}; generally speaking, when the Recorder received a new event it first checked to see if any of the involved objects were already present in the graph, created them if they are not, and then added a new relationship between the objects.
Visual examples of how provenance graphs were represented by the recorder follow in Section \ref{sec:casestudies}. 

\subsection{Network Guard}

Like the Database Capture Agent, the Network Guard is a multithreaded TCP proxy server that is placed between the web server and the network. When outbound traffic is transmitted from the web server, the Network Guard issues an ancestry query request to the Provenance Recorder by using the pid of the sending worker as a unique identifier.
Upon receipt of the list of ancestors, the Network Guard verifies policy compliance prior to permitting the data to be transmitted over the network.

\section{Evaluation}
\label{sec:eval}
\begin{table}[t]
\scriptsize
\centering
\begin{tabular}{ | l | c | }
\hline
{\bf Scenario} & {\bf Time} \\ \hline \hline
Web Application w/o \Sys & 34.5 ms\\ \hline
Web Application w/~ \Sys & 29.3 ms \\ \hline \hline
{\bf Overhead} & 5.1 ms (17.1\%) \\ \hline
\end{tabular}
\caption{End-to-end delay imposed by \Sys during the DVDStore Benchmark}
\label{tab:e2e}
\end{table}

\begin{table}[t]
\scriptsize
\centering
\begin{tabular}{ | l | l | c | }
\hline
{\bf Location} & {\bf Operation} & {\bf Time} \\ \hline \hline
Apache 2 {\tt config.c} & Transmit {\tt unit\_start} message & 0.82 ms\\ \hline
Apache 2 {\tt config.c} & Transmit {\tt unit\_end} message & 0.91 ms\\ \hline
Database Capture Agent & Parse SQL Query & 0.11 ms \\ \hline
Database Capture Agent & Transmit SQL Provenance & 1.98 ms \\ \hline
Database Capture Agent & Other (incl. proxy cost) & 1.28 ms\\ \hline
\end{tabular}
\caption{Microbenchmarks for \Sys system during the DVDStore benchmark}
\label{tab:microbench}
\end{table}

We evaluated \Sys using a VMWare Fusion VM running CentOS 6.5. We executed our webserver tests in the common deployment model of a virtual machine with 4GB RAM and 2 vCPUs. The host was a local server with two 2.4 GHz Quad-Core Intel Xeon processors and 12 GB RAM.

\begin{figure}[t]
\centering
\includegraphics[width=1\linewidth]{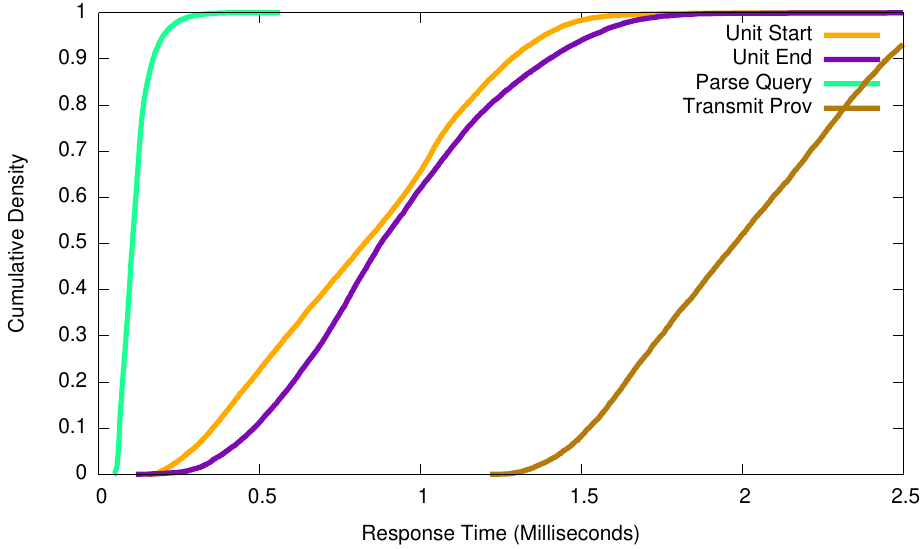}
\caption{Cumulative Densities of each microbenchmark step during the DVDStore benchmark.}
\label{fig:microbenchmark_cdfs}
\end{figure}

\pagebreak
\subsection{End-to-End Delay}

One of the vital measures of \Sys's performance is the end to end delay it imposes on web requests. The apparatus we used for both this test and its subsequent microbenchmarks was the Dell DVD Store Database Test Suite,\footnote{See \url{http://linux.dell.com/dvdstore/}} an open source simulation of an online ecommerce site. We configured DVDStore to run on MySQL with a 10 GB database. DVDStore's benchmarks are issued directly to the DBMS, bypassing the web front-end of the ecommerce site; in order to measure our modifications to Apache 2, we modified DVDStore's client workload driver to issue all SQL queries as curl requests to port 80 of the host. The queries were received by a toy PHP application on Apache 2 that relayed SQL queries to MySQL using POST requests, then returned the results to the client. In total, the DVDStore workload issued over 10,000 unique SQL queries to our system.

We measured overall performance under two configurations. In the first (without \Sys), an unmodified copy of httpd communicated directly with MySQL. In the second (with \Sys), our modified httpd communicated through the database capture agent. Table~\ref{tab:e2e} summarizes our findings. The average response time for queries when \Sys was disabled was 29.3 ms. The average response time for queries when \Sys was enabled was 34.5 ms, representing a cost of just \Cost, or \Overhead overhead.

\subsection{Microbenchmarks}

During the above trial, we instrumented \Sys to measure the time spent on individual steps involved in provenance capture -- 
	the {\tt unit\_start} and {\tt unit\_stop} messages generated by the web server, 
	the SQL parsing step, 
	and the transmission of provenance from the Database Capture Agent to the Provenance Recorder. 
By subtracting these measures from an end-to-end measurement, we also captured the approximate cost of other steps, most notably the delay imposed by proxying database traffic. 
The results are shown in Table \ref{tab:microbench}, and Figure \ref{fig:microbenchmark_cdfs} shows the associated cumulative density functions for each measure. 
The various SQL queries generated by DVDStore could be parsed and have their provenance extracted in an average of 0.11 ms. 
The primary source of delay in our \Sys system is due to inter-process communication; transmitting small provenance events to the recorder required approximately 1 ms, and larger messages required  approximately 2 ms.
Steps that required inter-process communication experienced high variance, indicating processing delays at the Recorder that could be addressed to improve performance.
As our provenance recorder implementation was single-threaded, it is likely that delays could be dramatically reduced through creating a multi-threaded version of the recorder.

\begin{figure}[t]
\centering
\includegraphics[width=1\linewidth]{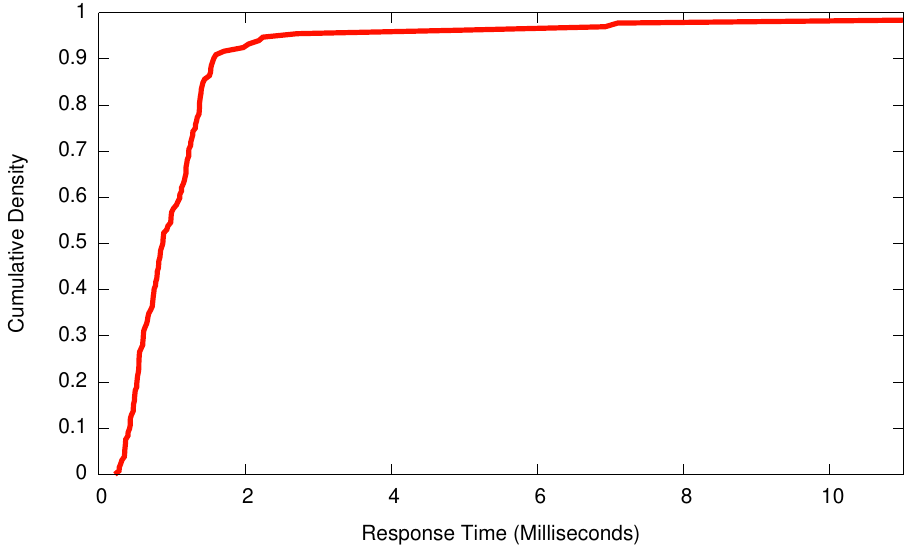}
\caption{Cumulative Density of Provenance Recorder's response time when queried by the Network Guard.}
\label{fig:query}
\end{figure}

\subsection{Network Guard Performance}
\label{sub:eval_query}

Benchmarking our Network Guard involved determining the speed with which \Sys provenance can be analyzed in a live system.
This was the first evaluation trial in which the Network Guard was enabled.
We instrumented the Network Guard to measure the time required to query the Provenance Recorder and process its response message. 
For this trial we repeated the DVDStore benchmark, which included a great diversity of SQL queries.
The results are shown in Figure \ref{fig:query}. The average response time by the Provenance Recorder was just 1.23 ms. We microbenchmarked this result as well, and found that on average 1.17 ms of this delay was due to IPC, while just 0.5 ms was required to generate the provenance ancestry. In the worst case, the query took 7 ms to respond, but this was also due to IPC delays and not to the cost of graph traversal. These results indicate that even our proof-of-concept implementation can be used as an enforcement mechanism without imposing unacceptable latency.

\subsection{Storage Overhead \& Optimization}

\begin{figure}[t]
\centering
\includegraphics[width=1\linewidth]{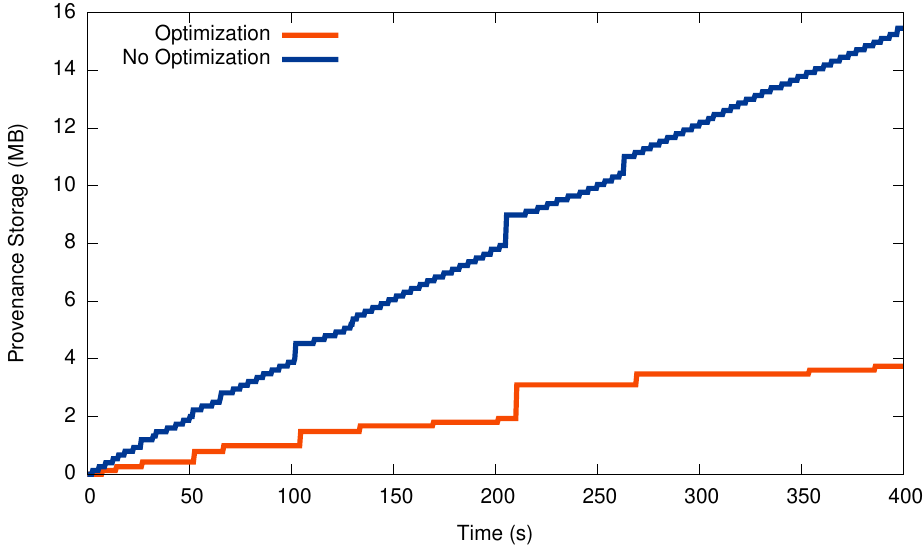}
\caption{Growth of provenance storage during the DVDStore benchmark}
\label{fig:storage}
\end{figure}

A vital consideration when collecting data provenance is the storage overhead incurred. Not only do high overheads increase the cost of storage, but preserving unnecessary provenance impacts the speed with which the provenance can be queried. To capture the storage overhead of \Sys, we observed the growth of the provenance graph during the DVDStore workload. Every 500 ms, we polled the total size of the provenance recorder's memory allocation using the {\tt /proc} file system.
The results of this trial are shown in Figure \ref{fig:storage}. 
During ust the first 6 minutes of the DVDStore trial, the provenance logs grew to 15.7 MB in size.
Unfortunately, for popular web services like Facebook and Twitter that field billions of requests per day, this would represent petabytes of provenance, making \Sys too demanding for widespread use.

To address this problem, we consider a garbage collection technique based on the SQLi defense scenario that was introduced in Section \ref{sub:networkguard}. In this circumstance, the vast majority of provenance will represent benign web requests. As a result, after the request is whitelisted by the Network Guard mechanism, this information can be discarded. In turn, those requests that were potentially malicious can be written to secondary storage for forensic analysis. By applying this technique, only the provenance of active requests needs be stored in memory. We implemented this garbage collection procedure in our provenance guard, and then repeated the trial. Figure \ref{fig:storage} shows that, after the optimization, the provenance store experienced logarithmic, rather than linear, growth. After 6 minutes, the provenance logs grew to just 3.9 MB in size, representing a 75\% decrease in storage overhead. While not a total solution to provenance storage overhead, this result indicates that overheads can grow manageably with the size of the web service.

\begin{figure*}[t]
\centering
\includegraphics[width=1\linewidth]{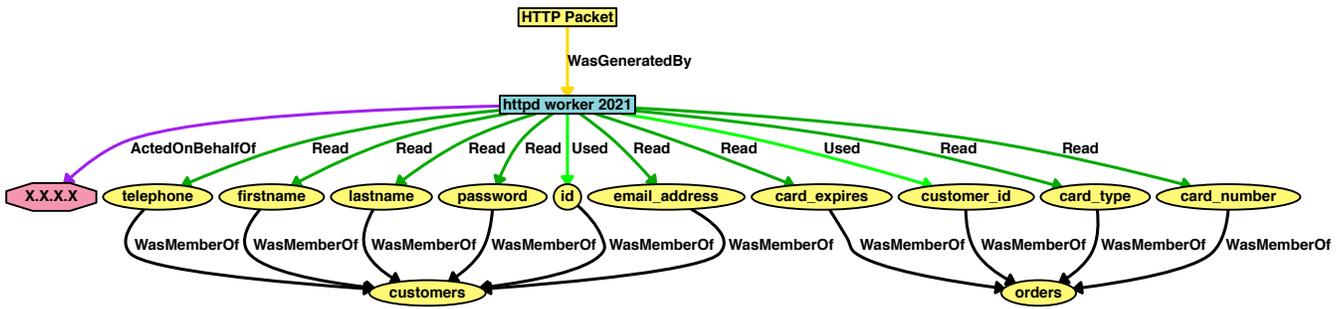}
\caption{The provenance of a successful SQL injection attack on an eCommerce site launched by remote host X.X.X.X. The attack exfiltrates several valuable data objects from the customers and orders tables.}
\label{fig:sqli_card_example}
\end{figure*}


\section{Case Studies}
\label{sec:casestudies}

In this section, we consider several attack scenarios in which \Sys can be used to monitor and prevent  Internet-based attacks.

\subsection{Scenario \#1: SQL Injection}
\label{sub:data_exfil}

Through the introduction of the Network Guard component, \Sys is able to prevent  SQLi-based data exfiltration attacks in real-time by performing provenance-based whitelisting of outbound server responses. As shown in 
Section \ref{sub:eval_query}, the Network Guard can authorize (or deny) an outbound message in just a few milliseconds. Here, we procedurally generate the provenance graph of a SQLi attack by using a toy PHP application on Apache 2 that relayed SQL queries to MySQL over POST requests. One example provenance ancestry is shown in Figure \ref{fig:sqli_card_example}. Here, we see the provenance of a message derived from SQL data from the customers and orders tables. Any number of the ancestral objects may be an indicator of a data leak. For example, it is unlikely that the web service would explicitly return passwords to the customer. Additionally, a web service would not return a full credit card number to the customer.

Interestingly, the query obfuscations that are commonly associated with SQLi are not present in the provenance graph. This is because such obfuscations are designed to bypass input sanitization checks that are performed by the web application. When a malicious input is able to successfully pass through these checks, the output is a well-formed SQL query. As a result, \Sys is optimally positioned to understand the intent of the attacker.



\begin{figure}[t]
\centering
\includegraphics[width=.9\linewidth]{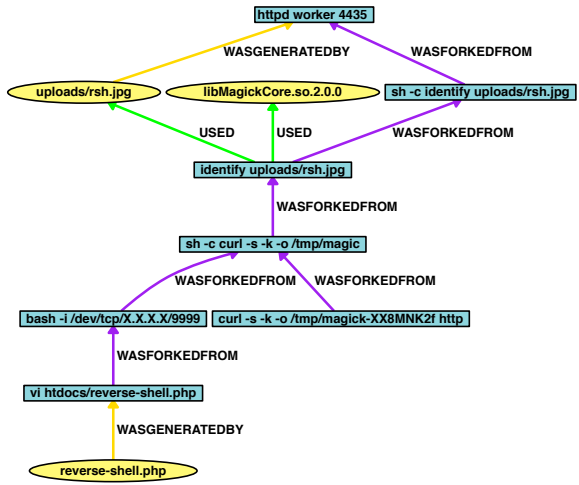}
\caption{The provenance of an ImageTragick exploit. Network activity has been pruned for clarity. The attacker uploads a malicious image file {\tt rsh.jpg} that opens a remote shell back to the attackers host. The attacker gains persistence on the machine by placing a reverse shell script in the {\tt htdocs} folder of the server.}
\label{fig:imagetragick}
\end{figure}

\subsection{Scenario \#2: ImageTragick Exploit}

To demonstrate the combined capabilities of \Sys and LPM when deployed in tandem, we developed a web application exploit based on the recently discovered vulnerabilities in the ImageMagick image processing library.\footnote{See https://imagetragick.com/} Our vulnerable web application made use of a PHP script that used the ImageMagick library to test whether an uploaded file was an image. The attack payload was a {\tt jpg} comprised of 4 lines of text including the following command:

\begin{center}
\tt image over 0,0 0,0 'https://127.0.0.1/x.php?x=`bash -i >\& /dev/tcp/X.X.X.X/9999 0>\&1`'
\end{center}

This code is executed by the server when the image is processed by the `identify` ImageMagick tool, causing a bash shell to be linked to the attackers remote host on port 9999. The provenance of these activities on the server side is shown in Figure \ref{fig:imagetragick}. Without \Sys, the operating system provenance for these events would be difficult to interpret. Dependency explosion would make it hard to identify which remote host was able to invoke a shell command. \Sys signals the start of a unit of work before the request is handled, which removes from suspicion all sessions that occurred prior to the compromise. Following the compromise, LPM can be securely configured such that the attackers actions on the system can continue to be monitored \cite{btb+2015}.


\begin{figure}[t]
\centering
\includegraphics[width=.9\linewidth]{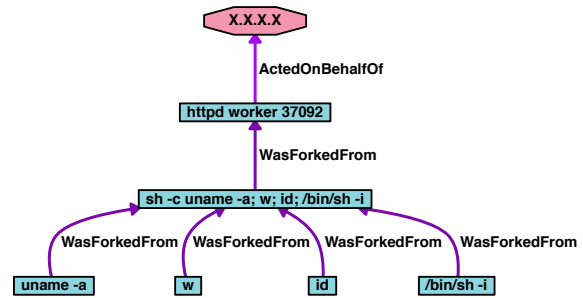}
\caption{The union of \Sys and systems layer provenance can track a remote shell invocation launched by remote host X.X.X.X. The provenance of file and packet manipulations have been pruned for clarity. \Sys assists the forensic process by identifying the remote host and unit of work responsible for the exploit.}
\label{fig:reverse_shell}
\end{figure}

\subsection{Scenario \#3: Reverse Shell Invocation}

As a final example, we monitor the attacker's subsequent visits to the web server through invoking the {\tt reverse-shell.php} script. 
To create realistic attack provenance, we made use of a publicly available php-reverse-shell application.\footnote{Available at \url{http://pentestmonkey.net/tools/web-shells/php-reverse-shell}.} 
Reverse shells were also an aspect of the Apache 2 Darkleech attack \cite{darkleech}. 
Figure \ref{fig:reverse_shell} shows the resulting provenance graph. Here, we can see that an {\tt httpd} worker with pid 37092 is handling a request from remote host X.X.X.X. 
Unexpectedly, the worker issues a serious of conspicuous system commands that collect information regarding the name of the machine ({\tt uname}), the identity of the active user ({\tt id}) which in this case is daemon, and the activities of other users that are currently logged in ({\tt w}). The worker then drops to shell.
This information would serve as an invaluable explanation as to the attackers intent once their intrusion had been discovered.

\section{Discussion}
\label{sec:disc}
\noindent
{\it Is \Sys a taint tracking system?}

Data provenance and taint analysis are two means of reasoning about information flow. 
While use of these techniques varies throughout the literature, one means of differentiating between the two is that taint analysis generally provides {\it what-provenance}, but not {\it how-}{\it provenance}. In a taint analysis solution for SQL Injection, a bit might be flipped in the taint label when a packet contained data from a particular table or column. While this would be sufficient to block the transmission of this packet, it would not provide forensic information about the attack. By providing both {\it what-} and {\it how-} provenance, our approach offers a concise explanation of the attack, including identifying the perpetrator.

\noindent
{\it Does \Sys track individual record accesses and updates?}

The finest granularity that our approach can offer is the granularity of columns; it cannot speak to the specific records that were impacted by a query. This limitation is comparable to the manner in which provenance and audit services track file manipulations at the system call layer; for instance, although the event of a process writing to a file is tracked, the log will not state what was written, or which lines of the file were changed. This choice of granularity represents a design tradeoff between performance and expressivity, and in the design of \Sys we have elected to follow convention. We show in Section \ref{sec:analysis} that this granularity can be effective at tracking attacks on web services. Moreover, in order to provide an efficient and rough estimation the impact of a query, we inspect the database engine's response to the query and measure either its size (in the case of a SELECT) or the number of records affected (in the case of INSERTs and UPDATEs).


\noindent
{\it What about NoSQL databases?}

Alternatives to tabular relational databases, including Accumulo, MongoDB, and Cassandra, have become increasingly popular due to their scalability and high performance. While our design focuses on relational databases, we feel that our approach is general enough to apply to these emergent technologies. Rather than extract table column names from queries, \Sys for Key-Value stores would instead focus on extracting Key names from queries. It is also worth noting that, in many cases, these systems already provide SQL-like query support. 

\section{Related Work}
\label{sec:relwork}

Our work is part of a growing body of literature that explores the use of provenance to address critical security challenges. Provenance has been employed to detect compromised nodes in data centers \cite{bbh+2014,gbm+2010,tbg+2011,zfn+2011}, explain and prevent data exfiltration \cite{btb+2015,jsd+2011}, and enrich access controls \cite{bmv+13,pds12}.

The notion of provenance tracking originated in literature from the database and scientific workflow communities. Systems such as Chimera offer management of manual provenance annotations \cite{fvw+2002}, but does not perform automatic collection.
The Kepler system offers automatic provenance recording for scientific workflows \cite{abj2006}, while VisTrails tracks provenance of data visualization procedures \cite{cfs+2006}.
Database management systems such as Trio \cite{w2004}, DBNotes \cite{ctv2005}, and ORCHESTRA \cite{kgi+2013} track the provenance of data records as they propagate through the database, and provide custom extensions to SQL so that provenance can be queried.
To reduce reliance on custom database engines, the PERM \cite{ga2009} and GProM \cite{agr+2014} systems perform automatic query rewriting to annotate result tuples with provenance information. 

In spite of the advances made in provenance-aware workflow engines, the systems are ``blind'' to the rest of the system. They cannot observe information flows beyond the boundaries of their own operation; critically, this means that they cannot make assertions about whether a given system object (e.g., network packet) was derived from a particular database record. This leaves them of little use when considering the SQL injection and data exfiltration scenarios that motivate our work. In contrast, our system assumes a ``black box" database engine, and observes SQL queries and results in order to provide efficient linkability between database operations and other system activity.

\subsection{Taint Analysis}

Like provenance, dynamic taint analysis tracks the propagation of data across a system. Automated instrumentation for taint tracking has been developed for x86 binaries \cite{ssp2008}, smartphones \cite{egc+2010}, and databases \cite{hom2006}, and dynamic taint analysis in sandbox environments has also been used to secure off-the-shelf applications \cite{zjj+2011}. Taint tracking
systems also suffer from the dependency explosion problem, an effect that can be mitigated, in part, by focusing on short-lived user-facing
applications like editors~\cite{fleming-icc12}.
Provenance can offer a more complete explanation as to {\it how} an object became tainted.  It is also more flexible: taint tracking relies on an immutable policy that requires that data be tagged at runtime, while a provenance-based approach can obtain a result after execution by  ``replaying''  the provenance graph \cite{zfn+2011}, permitting different taints to be considered without re-executing.

\subsection{Foundational Work}

The Nemesis system~\cite{Dalton-sec09} uses dynamic information flow tracking to prevent authentication bypass attacks.  Their system
requires manual annotation of the authentication table in the database and then performs taint analysis to track its use in the web
application.  In contrast, our system can prevent mis-use and exfiltration of any table in the database and does not require web 
application annotation.  Parno et al. built CLAMP~\cite{Parno-oakland09} to prevent exfiltration from typical web application servers.
CLAMP implicitly provides execution partitioning because they create an individual VM for each user's session.  They also provide
a database proxy, called the Query Restrictor, which filters queries to the database based on policy. 
In~\Sys, we collect the provenance of database accesses at the same location, but
deploy policy enforcement in a network guard after the web application has processed the request.
This allows us to specify high-level
policies that can potentially span multiple database accesses rather than needing to specify policy query by query.  Furthermore,
we also capture the provenance of the database response, which allows us to enforce policy on the number of records returned.

\section{Conclusion}
\label{sec:conc}
In spite of a pressing need for ways to explain and mitigate web application vulnerabilities, web services have received little attention as candidates for provenance capabilities. In this work, we presented \Sys, a system for creating provenance-aware web applications. Our system can be deployed without requiring any changes to the web application, yet provides rich, concise provenance graphs for web service workflows. We demonstrated \Sys's ability to explain, detect, and prevent SQL injection attacks, and to aid in the tracking of system layer attacks against the web server. In evaluation, we discovered that our system imposes just \Cost overhead on web requests, and introduced a optimization that dramatically reduces the storage burden of provenance capture. Thus, \Sys's non-invasive methodology demonstrates a deployment strategy for provenance not only in web services, but for all complex, heterogeneous application workflows.

\section*{Availability}
Our source code and testing framework will be released upon publication.

{
\footnotesize
\balance
\bibliographystyle{abbrv}
\bibliography{prov-lit-review,bates,sqli}

\begin{thebibliography}{10}

\bibitem{darkleech}
{Darkleech Apache Attacks Intensify}.
\newblock
  \url{http://www.darkreading.com/attacks-and-breaches/darkleech-apache-attacks-intensify/d/d-id/1109760?}

\bibitem{secondprovenancechallenge}
{The Second Provenance Challenge}.
\newblock
  \url{http://twiki.ipaw.info/bin/view/Challenge/SecondProvenanceChallenge}.

\bibitem{abj2006}
I.~Altintas, O.~Barney, and E.~Jaeger-Frank.
\newblock Provenance collection support in the kepler scientific workflow
  system.
\newblock In L.~Moreau and I.~Foster, editors, {\em Provenance and Annotation
  of Data}, volume 4145 of {\em Lecture Notes in Computer Science}, pages
  118--132. Springer Berlin Heidelberg, 2006.

\bibitem{agr+2014}
B.~Arab, D.~Gawlick, V.~Radhakrishnan, H.~Guo, and B.~Glavic.
\newblock {A Generic Provenance Middleware for Database Queries, Updates, and
  Transactions}.
\newblock June 2014.

\bibitem{bbh+2014}
A.~Bates, K.~Butler, A.~Haeberlen, M.~Sherr, and W.~Zhou.
\newblock {Let SDN Be Your Eyes: Secure Forensics in Data Center Networks}.
\newblock {SENT}, Feb. {2014}.

\bibitem{bmv+13}
A.~Bates, B.~Mood, M.~Valafar, and K.~Butler.
\newblock {Towards Secure Provenance-based Access Control in Cloud
  Environments}.
\newblock In {\em Proceedings of the 3rd ACM Conference on Data and Application
  Security and Privacy}, CODASPY '13, pages 277--284, New York, NY, USA, 2013.
  ACM.

\bibitem{btb+2015}
A.~Bates, D.~Tian, K.~R. Butler, and T.~Moyer.
\newblock {Trustworthy Whole-System Provenance for the Linux Kernel}.
\newblock In {\em Proceedings of 24th USENIX Security Symposium on USENIX
  Security Symposium}, Aug. 2015.

\bibitem{cfs+2006}
S.~P. Callahan, J.~Freire, E.~Santos, C.~E. Scheidegger, C.~T. Silva, and H.~T.
  Vo.
\newblock Vistrails: Visualization meets data management.
\newblock In {\em Proceedings of the 2006 ACM SIGMOD International Conference
  on Management of Data}, SIGMOD '06, pages 745--747, New York, NY, USA, 2006.
  ACM.

\bibitem{ctv2005}
L.~Chiticariu, W.-C. Tan, and G.~Vijayvargiya.
\newblock {DBNotes: A Post-it System for Relational Databases Based on
  Provenance}.
\newblock In {\em Proceedings of the 2005 ACM Special Interest Group on
  Management of Data Conference}, SIGMOD'05, June 2005.

\bibitem{Dalton-sec09}
M.~Dalton, C.~Kozyrakis, and N.~Zeldovich.
\newblock Nemesis: Preventing authentication \&\#38; access control
  vulnerabilities in web applications.
\newblock In {\em Proceedings of the 18th Conference on USENIX Security
  Symposium}, SSYM'09, pages 267--282, Berkeley, CA, USA, 2009. USENIX
  Association.

\bibitem{egc+2010}
W.~Enck, P.~Gilbert, B.-G. Chun, L.~P. Cox, J.~Jung, P.~McDaniel, and A.~N.
  Sheth.
\newblock {TaintDroid: An Information-flow Tracking System for Realtime Privacy
  Monitoring on Smartphones}.
\newblock In {\em Proceedings of the 9th USENIX Symposium on Operating Systems
  Design and Implementation}, OSDI'10, Oct. 2010.

\bibitem{fleming-icc12}
C.~Fleming, P.~Peterson, E.~Kline, and P.~Reiher.
\newblock Data tethers: Preventing information leakage by enforcing
  environmental data access policies.
\newblock In {\em Communications (ICC), 2012 IEEE International Conference on},
  pages 835--840, June 2012.

\bibitem{fvw+2002}
I.~T. Foster, J.-S. V\"{o}ckler, M.~Wilde, and Y.~Zhao.
\newblock {Chimera: AVirtual Data System for Representing, Querying, and
  Automating Data Derivation}.
\newblock In {\em Proceedings of the 14th Conference on Scientific and
  Statistical Database Management}, SSDBM'02, July 2002.

\bibitem{gbm+2010}
A.~Gehani, B.~Baig, S.~Mahmood, D.~Tariq, and F.~Zaffar.
\newblock {Fine-grained Tracking of Grid Infections}.
\newblock In {\em Proceedings of the 11th IEEE/ACM International Conference on
  Grid Computing}, GRID'10, Oct 2010.

\bibitem{gl2007}
A.~Gehani and U.~Lindqvist.
\newblock {Bonsai: Balanced Lineage Authentication}.
\newblock In {\em Proceedings of the 23rd Annual Computer Security Applications
  Conference}, ACSAC'07, Dec 2007.

\bibitem{gt2012}
A.~Gehani and D.~Tariq.
\newblock {SPADE: Support for Provenance Auditing in Distributed Environments}.
\newblock In {\em Proceedings of the 13th International Middleware Conference},
  Middleware '12, Dec 2012.

\bibitem{ga2009}
B.~Glavic and G.~Alonso.
\newblock {Perm: Processing Provenance and Data on the Same Data Model Through
  Query Rewriting}.
\newblock In {\em Proceedings of the 25th IEEE International Conference on Data
  Engineering}, ICDE '09, Mar. 2009.

\bibitem{gm2011}
P.~Groth and L.~Moreau.
\newblock {Representing Distributed Systems Using the Open Provenance Model}.
\newblock {\em Future Gener. Comput. Syst.}, 27(6):757--765, June 2011.

\bibitem{ho2005}
W.~G.~J. Halfond and A.~Orso.
\newblock Amnesia: Analysis and monitoring for neutralizing sql-injection
  attacks.
\newblock In {\em Proceedings of the 20th IEEE/ACM International Conference on
  Automated Software Engineering}, ASE '05, 2005.

\bibitem{hom2006}
W.~G.~J. Halfond, A.~Orso, and P.~Manolios.
\newblock Using positive tainting and syntax-aware evaluation to counter sql
  injection attacks.
\newblock In {\em Proceedings of the 14th ACM SIGSOFT International Symposium
  on Foundations of Software Engineering}, SIGSOFT '06/FSE-14, pages 175--185,
  New York, NY, USA, 2006. ACM.

\bibitem{hsw2009}
R.~Hasan, R.~Sion, and M.~Winslett.
\newblock {The Case of the Fake Picasso: Preventing History Forgery with Secure
  Provenance}.
\newblock In {\em Proceedings of the 7th USENIX Conference on File and Storage
  Technologies}, FAST'09, San Francisco, CA, USA, Feb. 2009.

\bibitem{jsd+2011}
S.~N. Jones, C.~R. Strong, D.~D.~E. Long, and E.~L. Miller.
\newblock {Tracking Emigrant Data via Transient Provenance}.
\newblock In {\em 3rd Workshop on the Theory and Practice of Provenance},
  TAPP'11, June 2011.

\bibitem{kgi+2013}
G.~Karvounarakis, T.~J. Green, Z.~G. Ives, and V.~Tannen.
\newblock Collaborative data sharing via update exchange and provenance.
\newblock {\em ACM Trans. Database Syst.}, 38(3):19:1--19:42, Sept. 2013.

\bibitem{lzx2013}
K.~H. Lee, X.~Zhang, and D.~Xu.
\newblock {High Accuracy Attack Provenance via Binary-based Execution
  Partition}.
\newblock In {\em Proceedings of the 20th ISOC Network and Distributed System
  Security Symposium}, NDSS'13, Feb.

\bibitem{Lee-ccs13}
K.~H. Lee, X.~Zhang, and D.~Xu.
\newblock Loggc: garbage collecting audit log.
\newblock In {\em Proceedings of the 2013 ACM SIGSAC conference on Computer
  \&\#38; communications security}, CCS '13, pages 1005--1016, New York, NY,
  USA, 2013. ACM.

\bibitem{lm2010}
J.~Lyle and A.~Martin.
\newblock {Trusted Computing and Provenance: Better Together}.
\newblock In {\em 2nd Workshop on the Theory and Practice of Provenance},
  TaPP'10, Feb. 2010.

\bibitem{mzx2016}
S.~Ma, X.~Zhang, and D.~Xu.
\newblock {ProTracer: Towards Practical Provenance Tracing by Alternating
  Between Logging and Tainting}.
\newblock In {\em Proceedings of the 23rd ISOC Network and Distributed System
  Security Symposium}, NDSS, Feb. 2016.

\bibitem{ms2012}
P.~Macko and M.~Seltzer.
\newblock {A General-purpose Provenance Library}.
\newblock In {\em 4th Workshop on the Theory and Practice of Provenance},
  TaPP'12, June 2012.

\bibitem{mkb+2009}
K.-K. Muniswamy-Reddy, U.~Braun, D.~A. Holland, P.~Macko, D.~Maclean, D.~Margo,
  M.~Seltzer, and R.~Smogor.
\newblock {Layering in Provenance Systems}.
\newblock In {\em Proceedings of the 2009 Conference on USENIX Annual Technical
  Conference}, ATC'09, June 2009.

\bibitem{mhb+2006}
K.-K. Muniswamy-Reddy, D.~A. Holland, U.~Braun, and M.~Seltzer.
\newblock {Provenance-aware Storage Systems}.
\newblock In {\em Proceedings of the Annual Conference on USENIX '06 Annual
  Technical Conference}, Proceedings of the 2006 Conference on USENIX Annual
  Technical Conference, June 2006.

\bibitem{nps12}
D.~Nguyen, J.~Park, and R.~Sandhu.
\newblock {Dependency Path Patterns As the Foundation of Access Control in
  Provenance-aware Systems}.
\newblock In {\em Proceedings of the 4th USENIX Conference on Theory and
  Practice of Provenance}, TaPP'12, pages 4--4, Berkeley, CA, USA, 2012. USENIX
  Association.

\bibitem{pds12}
J.~Park, D.~Nguyen, and R.~Sandhu.
\newblock {A Provenance-Based Access Control Model}.
\newblock In {\em Proceedings of the 10th Annual International Conference on
  Privacy, Security and Trust (PST)}, pages 137--144, 2012.

\bibitem{Parno-oakland09}
B.~Parno, J.~M. McCune, D.~Wendlandt, D.~G. Andersen, and A.~Perrig.
\newblock Clamp: Practical prevention of large-scale data leaks.
\newblock In {\em Proceedings of the 2009 30th IEEE Symposium on Security and
  Privacy}, SP '09, pages 154--169, Washington, DC, USA, 2009. IEEE Computer
  Society.

\bibitem{pmm+2012}
D.~Pohly, S.~McLaughlin, P.~McDaniel, and K.~Butler.
\newblock {Hi-Fi: Collecting High-Fidelity Whole-System Provenance}.
\newblock In {\em Proceedings of the 28th Annual Comptuer Security Applications
  Conference}, ACSAC'12, Orlando, FL, USA, 2012.

\bibitem{sc2005}
C.~Sar and P.~Cao.
\newblock {Lineage File System}.
\newblock http://crypto.stanford.edu/{\textasciitilde}cao/lineage.html.

\bibitem{ssp2008}
P.~Saxena, R.~Sekar, and V.~Puranik.
\newblock {Efficient Fine-grained Binary Instrumentationwith Applications to
  Taint-tracking}.
\newblock In {\em Proceedings of the 6th Annual IEEE/ACM International
  Symposium on Code Generation and Optimization}, CGO '08, pages 74--83, New
  York, NY, USA, 2008. ACM.

\bibitem{tbg+2011}
D.~Tariq, B.~Baig, A.~Gehani, S.~Mahmood, R.~Tahir, A.~Aqil, and F.~Zaffar.
\newblock {Identifying the Provenance of Correlated Anomalies}.
\newblock In {\em Proceedings of the 2011 ACM Symposium on Applied Computing},
  SAC '11, Mar. 2011.

\bibitem{w2004}
J.~Widom.
\newblock {Trio: A System for Integrated Management of Data, Accuracy, and
  Lineage}.
\newblock Technical Report 2004-40, Stanford {InfoLab}, Aug. 2004.

\bibitem{w3c}
{World Wide Web Consortium}.
\newblock {PROV-Overview: An Overview of the PROV Family of Documents}.
\newblock \url{http://www.w3.org/TR/prov-overview/}, 2013.

\bibitem{zpj+2006}
L.~Zhang, A.~Persaud, A.~Johnson, and Y.~Guan.
\newblock Detection of stepping stone attack under delay and chaff
  perturbations.
\newblock In {\em Performance, Computing, and Communications Conference, 2006.
  IPCCC 2006. 25th IEEE International}, pages 10 pp.--256, April 2006.

\bibitem{zfn+2011}
W.~Zhou, Q.~Fei, A.~Narayan, A.~Haeberlen, B.~T. Loo, and M.~Sherr.
\newblock {Secure Network Provenance}.
\newblock In {\em Proceedings of the 23rd ACM Symposium on Operating Systems
  Principles}, SOSP'11, Oct. 2011.

\bibitem{zjj+2011}
D.~Y. Zhu, J.~Jung, D.~Song, T.~Kohno, and D.~Wetherall.
\newblock {TaintEraser: Protecting Sensitive Data Leaks Using Application-level
  Taint Tracking}.
\newblock {\em SIGOPS Oper. Syst. Rev.}, 45(1):142--154, Feb. 2011.

\end{thebibliography}
}
\end{document}